\documentclass[showpacs,twocolumn,aps,prl]{revtex4}
\usepackage{amsmath,amssymb}
\begin{document}
\def\cs#1#2{#1_{\!{}_#2}}
\def\css#1#2#3{#1^{#2}_{\!{}_#3}}
\def\ket#1{|#1\rangle}
\def\bra#1{\langle#1|}

\title{Testing Gravity-Driven Collapse of the Wavefunction via Cosmogenic Neutrinos}

\author{Joy Christian}

\affiliation{Perimeter Institute, 31 Caroline Street North, Waterloo, Ontario N2L 2Y5, Canada,}
\affiliation{Department of Physics, University of Oxford, Parks Road, Oxford OX1 3PU, England}

\begin{abstract}
It is pointed out that the Di\'osi-Penrose ansatz for gravity-induced quantum state reduction
can be tested by observing oscillations in the flavor ratios of neutrinos originating at
cosmological distances. Since such a test would be almost free of environmental decoherence,
testing the ansatz by means of a next generation neutrino detector such as IceCube would be much
cleaner than by experiments proposed so far involving superpositions of macroscopic systems. The
proposed microscopic test would also examine the universality of superposition
principle at unprecedented cosmological scales.
\end{abstract}


\maketitle

In the 1950s Feynman observed that a possible gravity-induced failure of quantum mechanics for
objects as small as the Planck mass (${\sim 10^{-5}}$ grams) is not inconsistent with the
existing physical evidence \cite{Feynman-1957}. Remarkably, this elementary observation remains
unchallenged even today, when we have gained almost unshakable confidence in the universality
of quantum mechanics. Compounded by persistent conceptual problems of both quantum theory and
quantum gravity \cite{Christian-2000}, time and again this fact has inspired suggestions of
a gravity-driven reduction of the state vector \cite{reduction}\cite{Diosi}\cite{Penrose}, albeit
in a considerably evolved form than its preliminary inception by Feynman \cite{Feynman-1957}.

Noteworthy among these are the specific proposals put forward by Di\'osi \cite{Diosi} and
Penrose \cite{Penrose}, who independently arrive at a phenomenological ansatz for the time scale
beyond which quantum superpositions may become unstable. This ``duration of quantal stability''
turns out to be experimentally testable, and can be expressed in a form resembling the mean life
of an unstable particle:
\begin{equation}
{\cal T}\sim\frac{\hbar}{\,\Delta E^{\,\!\,}_G}\,,\label{first}
\end{equation}
where ${\Delta E^{\,\!\,}_G}$ is the gravity-induced ill-definedness in the energy of a given
system in superposition of two states.

There have been several experimental proposals to test this ansatz
\cite{Penrose}\cite{Pearle-1994}\cite{Marshall-2003}. Remarkably, one of these
proposals \cite{Marshall-2003} purports to superpose a mirror of some ${10^{-10}}$
grams, which is only about 5 orders of magnitude short of the Planck mass. These proposals
are, of course, only a small part of the ongoing drive to experimentally push the boundaries of
superposition principle as far up the macroscopic scale as possible \cite{macro-exp}. All of
these efforts are hampered, however, by one major difficulty.
Due to the intractability of environmentally induced decoherence for such large systems,
it is usually extremely difficult to distinguish any genuine state reduction scheme from the
effective decoherence resulting from a subjective omission of the environmental degrees of
freedom \cite{Joos-1985}.

There is, however, nothing in ansatz (\ref{first}) that necessitates a macroscopic system for
its validity. Indeed, if the ansatz was meant only for such large systems, it would not
bear the fundamental and universal significance attached to it by Di\'osi and Penrose. 
Put differently, even if the ansatz is verified for macroscopic systems, it cannot be accepted
as a truly fundamental feature of the world until it is also verified for elementary systems.
Therefore, here we propose to test the ansatz on neutrinos, originating at cosmological
distances. Since neutrinos are electromagnetically neutral and sensitive only to the weak and
gravitational interactions, the chances of them decohering within the cosmic vacuum are
negligible. As a result, cosmogenic neutrinos form an ideal system for testing any scheme of
gravity-driven state reduction.

To appreciate this, in what follows we first take a closer look at the rational behind ansatz
(\ref{first}), then review the theory of neutrino flavor oscillations, and, finally, extract
deviations from the quantum mechanically expected flavor ratios by applying the ansatz to
massive neutrinos.

{\em The physics of Di\'osi-Penrose ansatz}.---Since Penrose's proposal is minimalist
in conception (i.e., it relies only on the first principles of quantum mechanics and general
relativity), we shall follow his reasoning \cite{Penrose}. He considers two quantum states
of a given mass,
${\ket{\Psi_1}}$ and ${\ket{\Psi_2}}$, each {\it stationary} on its own, and possessing the
same energy $E$:
\begin{equation}
i\,\frac{\partial}{\partial t}\,
\ket{\Psi_1}\;=\;E\,\ket{\Psi_1}\;,\;\;\;\;\;\;\;
i\,\frac{\partial}{\partial t}\,
\ket{\Psi_2}\;=\;E\,\ket{\Psi_2}
\end{equation}
(henceforth we mostly use Planck units: ${\hbar=c=G=1}$).
In standard quantum mechanics linearity necessitates that a possible superposition of these two
states, such as ${\ket{{\cal X}}=\lambda_1\ket{\Psi_1}+\lambda_2\ket{\Psi_2}}$,
must itself be a stationary state with the same energy. However, when gravitational fields of
the two masses are taken into account, in general each of the original states would
correspond to two entirely {\it different} spacetimes. The principles of general relativity
would then dictate that the time-translation operators
`${\frac{\partial\,}{\partial\,t}}$', corresponding to the action of the timelike Killing
vector fields of the two (stationary) spacetimes, would also be quite distinct form
one another. On the other hand, when these two Killing fields
happen not to be too different, there would be only a slight ill-definedness in the action of
${\frac{\partial\,}{\partial\,t}}$, and that would be reflected in the energy of the
system. Penrose uses this gravity-induced ``error'' in energy, ${\Delta E^{\!\,\!}_G}$,
as an approximate measure of instability of the superposition, and postulates
the mean lifetime ${\cal T}$ of such a state to be ${[\Delta E^{\!\,\!}_G]^{-1}}$, as in
(\ref{first}), with two decay modes being the component states ${\ket{\Psi_1}}$ and
${\ket{\Psi_2}}$ with probabilities ${|\lambda_1|^2}$ and ${|\lambda_2|^2}$, respectively.

Penrose next suggests that this measure of instability, ${\Delta E^{\!\,\!}_G}$, can be
estimated in terms of the incompatibility between the notions of free fall within the two
spacetimes. At some identifiable event, let ${{\bf a}_1}$ and ${{\bf a}_2}$ be the acceleration
3-vectors of the free-fall motions in the two respective spacetimes. Then
${\Delta E^{\!\,\!}_G}$ can be estimated as
\begin{equation}
\Delta E^{\!\,\!}_G \approx \,\xi\int_{\Sigma_t}
\left({\bf a}_1-{\bf a}_2\right)\cdot\left({\bf a}_1-{\bf a}_2\right)\;
d{\bf r}\,,\label{pre-second}
\end{equation}
where the integrand is a coordinate-independent scalar quantity, ${\Sigma_t}$ represents
a three-dimensional hypersurface at an instant of time ${t}$, and ${\xi\geq 0}$ is an arbitrary
dimensionless
parameter (in what follows, this parameter will provide a phenomenological handle on the
``strength'' of quantum state reduction). Of course, in Newtonian approximation ${{\bf a}_1}$
and ${{\bf a}_2}$ are simply the forces per unit test mass: ${{{\bf a}_1}=-\nabla\Phi_1}$ and
${{{\bf a}_2}=-\nabla\Phi_2}$, where ${\Phi_1}$ and ${\Phi_2}$ are the respective gravitational
potentials for the two spacetimes. Therefore, using the Poisson's equation
${\nabla^2\Phi({\bf r})=4\pi\,\rho({\bf r})}$, the estimate (\ref{pre-second}) can be reduced to
\begin{equation}
\Delta E^{\!\,\!}_G \!\approx 4\pi\xi\!\int\!\!\!\int
\frac{\left[\rho_1\!\left({\bf r}\right)-\rho_2\!\left({\bf r}\right)\right]\,
\left[\rho_1\!\left({\bf r'}\right)-\rho_2\!\left({\bf r'}\right)\right]}{|{\bf r}-{\bf r'}|}\;
d{\bf r}d{\bf r'}\!,\label{second}
\end{equation}
where ${\rho_1}$ and ${\rho_2}$ are the two respective mass distributions responsible for the
two spacetimes \cite{Penrose}. It is worth noting here that this ill-definedness in energy is
based on the gravitational energy of the system itself, and not on the energy of any externally
present fields, although the latter may play an indirect role in some cases. In fact, it is
essentially the gravitational self-energy of the {\it difference} between the two superposed
mass distributions.

The order of magnitude for the mean life ${\cal T}$ based on the expression (\ref{second}) can
now be estimated to be simply ${\Delta r/m^2}$, where ${m}$ is the rest mass of the system, and
${\Delta r}$ is the spread in the position of the system between its two superposed states. For
example, in the case of a nucleon, with ${\Delta r}$ taken to be its strong interaction
range, the mean life of a superposition of its states turns out to be over ${10^7}$ years;
whereas for systems as large as a speck of dust of mass ${10^{-4}}$ grams and position spread
1 mm, it plunges to some ${10^{-13}}$ seconds (cf. \cite{Diosi}\cite{Penrose}). Evidently,
the postulated mean life ${\cal T}$ reproduces the phenomenology of quantum state reduction
quite compellingly.

Despite this predicted astronomically long mean life of superpositions for elementary
particles, the proposed ansatz turns out to be testable for cosmogenic neutrinos.

{\em Theory of neutrino oscillations}.---The remarkable phenomena of neutrino
oscillations are due to the fact that neutrinos of definite flavor states
${\ket{\nu_{\alpha}}}$, ${\alpha=e,\mu,}$ or ${\tau}$, are {\it not} particles of definite mass
states ${\ket{{\nu_j}}}$, ${j=1,2,}$ or ${3}$, but are superpositions of the definite mass
states \cite{Kayser-2001}:
\begin{equation}
\ket{\nu_{\alpha}}=\sum_j U^*_{\alpha j}\,\ket{\nu_j},\label{super}
\end{equation}
with ${U}$ being the (time-independent) leptonic mixing matrix. By the same token, neutrinos of
definite mass states are superpositions of the definite flavor states:
${\ket{\nu_j}=\sum_{\beta} U_{\beta j}\,\ket{\nu_{\beta}}}$,
with the mixing matrix being subject to the unitarity constraint
${\sum_j U^*_{\alpha j} U_{\beta j}=\delta_{\alpha\beta}}$. As a neutrino of
definite flavor state propagates through vacuum for a long enough
laboratory time, the heavier mass-eigenstates in (\ref{super}) lag behind the lighter ones, and
the neutrino transforms itself into a different flavor state.
The probability for this transition from one flavor
state to another can be easily obtained as follows. In the rest frame of each
${\ket{\nu_j}}$, where the proper time is ${\tau_j}$, plane wave analysis leads to the
Schr\"odinger equation
\begin{equation}
i\frac{\partial\;}{\partial\tau_j}\ket{\nu_j(\tau_j)}=m_j\,\ket{\nu_j(\tau_j)},\label{dynamics}
\end{equation} 
with a solution ${\ket{\nu_j(\tau_j)}=e^{-im_j\tau_j}\ket{\nu_j(0)}}$, where ${m_j}$ is the
eigenvalue of the mass-eigenstate ${\ket{\nu_j(0)}}$. In terms of the coordinate time ${t}$
and position ${\bf r}$ in the laboratory frame, this phase factor takes the familiar form
\begin{equation}
e^{-i(E_jt-{\bf p}_j\cdot{\bf r})},\label{phase}
\end{equation}
where ${E_j}$ and ${{\bf p}_j}$ are, respectively, the energy and momentum associated with
the definite mass state ${\ket{\nu_j(0)}}$.

Now neutrinos are highly relativistic particles, which permits the assumption that
${t\approx |{\bf r}|}$ ${=L}$, where ${L}$ is the distance traveled by them before detection.
Moreover, assuming that they are produced with the same energy ${E}$ regardless of which state
${\ket{\nu_j(0)}}$ they are in (and that ${m_j\ll E}$), up to the second order in ${m_j}$ the
dispersion relation gives the following expression for their momenta,
\begin{equation}
p_j=\sqrt{E^2-m_j^2}\,\approx E-\frac{\,m^2_j\,}{2E\,},\label{momen}
\end{equation}
which, along with the assumption ${t\approx L}$, reduces the phase factor in (\ref{phase}) to
${e^{-i(m^2_j/2E)L}}$. Consequently, in the laboratory frame, and up to the second
order in ${m_j}$, the time evolution of the neutrino flavor state (\ref{super}) is given by
\begin{equation}
\ket{\nu_{\alpha}(t)}
=\sum_{\beta}\sum_j U^*_{\alpha j}\,e^{-i(m^2_j/2E)L}\,U_{\beta j}\,
\ket{\nu_{\beta}(0)}.\label{super-time}
\end{equation}
As a result,
the transition probability for the neutrinos to ``oscillate'' from a given flavor state, say
${\ket{\nu_{\alpha}(0)}}$, to another flavor state, say ${\ket{\nu_{\beta}(t)}}$, is given by
\begin{equation}
\begin{split}
P_{\alpha\beta}(E,\,L):=&\;P_{\nu_{\alpha}\rightarrow\,\nu_{\beta}}
(E,\,L)=|\langle\nu_{\beta}(0)\ket{\nu_{\alpha}(t)}|^2 \\ \label{transi-prob}
=\delta_{\alpha\beta}&-\!\sum_{j\not=k}U^*_{\alpha j}U_{\alpha k}U_{\beta j}U^*_{\beta k}
\!\left[1-e^{-i(\Delta m^2_{jk}/2E)L}\right]\!, \\
\end{split}
\end{equation}
where ${\Delta m_{jk}^2}$ ${\equiv}$ ${m_k^2-m_j^2}$ ${> 0}$ is the difference in the squares of
the two masses. From this transition probability it is clear that the experimental observability
of neutrino oscillations is determined by the quantum phase
\begin{equation}
\Phi:=2\pi\,\frac{L\,}{L_O}\,,\label{quan-phase}
\end{equation}
where ${L_O(E,\,m):=4\pi E/\Delta m_{jk}^2}$ is the energy-dependent oscillation length. Thus,
flavor changes would be observable whenever the propagation distance ${L}$ is of the order of the
oscillation length ${L_O}$. Therefore, in what follows it would suffice to concentrate on these
two variables.

{\em Applying Di\'osi-Penrose ansatz to massive neutrinos}.---It is evident from ansatz
(\ref{first}) that the proposed mean life of superpositions is independent of the speed of
light, and hence applicable to both nonrelativistic as well as relativistic systems, including
ultra-relativistic neutrinos. Moreover, for our purposes it would not be incongruous to
estimate
the spacetime distortions due to neutrinos themselves by treating them as classical spinning
particles. Accordingly, let us consider a spherically symmetric gravitating body of mass ${m}$
and angular momentum ${\bf s}$. If the gravitational field produced by the body is sufficiently
weak, then, in an approximate global inertial frame, it can be described by the following
well-known solution of the linearized Einstein's field equations:
\begin{equation}
\begin{split}
\!\!ds^2 \approx\,&-\left(1-\frac{2m}{r}\right)dt^2-\frac{4|{\bf s}|}{r}
\sin^2\theta \;dt\,d\phi \\
&+\left(1+\frac{2m}{r}\right)\left(dr^2+r^2d\theta^2+r^2\sin^2\theta\,d\phi^2\right).
\label{metric} \\
\end{split}
\end{equation}
This is essentially a Newtonian line element, apart from the off-diagonal term involving the
magnitude ${|{\bf s}|}$ of the intrinsic angular momentum of the body. Now, in the case of a
neutrino the magnitude of ${\bf s}$ is simply ${\frac{\hbar}{2}}$, which, in ordinary units,
is some 16 orders of magnitude per second smaller than the estimated (active) neutrino
mass (measured to be ${< 2.3}$ eV \cite{Kraus-2004}). Therefore, it would be adequate for our
purposes to consider only the Newtonian part of the gravitational field due to the neutrino
mass, and neglect the off-diagonal contribution due to its spin.

As an excellent approximation, it is then possible to apply Penrose's Newtonian prescription
(\ref{second}) to each of the three pairs of neutrino states in the superposition (\ref{super}).
Moreover, provided we continue to take the expectation value for the neutrino mass-distribution
to be a uniform sphere of effective radius ${a_j}$, this two-body Newtonian prescription can be
easily calculated to be
\begin{equation}
\Delta E^{j,k}_G \approx 8\pi\xi\left[\frac{3\,m_j^2}{5\,a_j}+\frac{3\,m_k^2}{5\,a_k}-
\frac{m_j\,m_k}{|{\bf r}_j-{\bf r}_k|}\right],\label{theifor}
\end{equation}
with ${|{\bf r}_j-{\bf r}_k|}$ being the displacement between the two superposed mass-eigenstates
resulting from their journey. The particular shape of the smearing introduced here to avoid the
self-energy divergence has little effect on what follows \cite{Ghirardi-1990}. More importantly,
it is manifest from (\ref{theifor}) that the product ${\Delta E^{j,k}_G\times dL}$ is Lorentz
invariant. Now for ultra-relativistic neutrinos the usual spreading of the wavepacket can be
easily shown to be negligible \cite{Bottino-1989}, but within a neutrino beam of definite energy
the different mass-eigenstates ${\ket{\nu_j}}$ in the superposition (\ref{super}) travel at
slightly different speeds ${\beta_j}$, producing the displacement
\begin{equation}
|{\bf r}_j-{\bf r}_k|=(\beta_j-\beta_k)t\approx\frac{(p_j-p_k)}{E}L\approx
\frac{\Delta m^2_{jk}}{2E^2}L\,.\label{giving}
\end{equation}
Here the last relation follows from (\ref{momen}), and we have used the relativistic identity
${\beta=p/E}$, as well as continued to assume ${t\approx L}$ and taken the ${k^{th}}$ neutrino
to be the heavier (and hence the slower) of the two partners.
Using (\ref{giving}), the measure (\ref{theifor}) can now be rewritten in terms of neutrino
parameters---such as energy, propagation length, and the mass-squared difference---as follows:
\begin{equation}
\Delta E^{j,k}_G(L) \approx 8\pi\xi
\left[\frac{3\left(m_j+m_k\right)}{5\,G^{\,}_{\!F}}-
\frac{\,2\,m_j\,m_k\,E^2}{\Delta m^2_{jk}\,L}\right],\label{diosi-pen}
\end{equation}
where we have taken the effective radii ${a_j}$ to be ${\approx G^{\,}_{\!F}\,m_j}$, with
${G^{\,}_{\!F}}$ being the Fermi constant of weak interactions.

The effect of this gravity-induced ill-definedness on the off-diagonal matrix elements of the
statistical operator corresponding to the superposition (\ref{super-time}) can now be easily
worked out (cf. \cite{Ghirardi-1990}). Unsurprisingly, it turns out to be a
{\it time-dependent} modification of the matrix elements,
\begin{equation}
U^*_{\alpha j}U_{\alpha k}U_{\beta j}U^*_{\beta k}\longrightarrow
e^{-\left[\int_{D}^L\Delta E^{j,k}_G(L')\,dL'\right]}
U^*_{\alpha j}U_{\alpha k}U_{\beta j}U^*_{\beta k}\,,\label{time-modi}
\end{equation}
where the integrand---with definition ${\Delta E^{j,k}_G(D)\equiv0}$---is the ``decay
constant'' corresponding to the ``mean life'' ${\cal T}$ in (\ref{first}). As a result
of this non-unitary modification, the transition
probability (\ref{transi-prob}) for flavor oscillations would acquire a time-dependent
``damping factor'':
\begin{equation}
\begin{split}
P_{\alpha\beta}(E,\,L)\longrightarrow\;&\delta_{\alpha\beta}-\!\sum_{j\not=k}\,
U^*_{\alpha j}U_{\alpha k}U_{\beta j}U^*_{\beta k} \\ \label{modi-prob}
&\;\;\times\left[1-e^{-i(\Delta m^2_{jk}/2E)L\,-\int_{D}^L\Delta E^{j,k}_G(L')\,dL'}\right]\!.\\
\end{split}
\end{equation}
Now, in the absence of the damping factor (i.e., within a unitary mechanics), it is clear from
the phase (\ref{quan-phase}) that flavor changes can be observable only when the propagation
distance ${L}$ of neutrinos is about the same size as their oscillation length ${L_O}$---i.e.,
only when the condition
\begin{equation}
\Delta m_{jk}^2\sim\frac{4\pi E}{L}\label{obscon}
\end{equation}
is satisfied. Substituting this observability condition into the evaluation of the 
equation ${e^{-\left[\int_{D}^L\Delta E^{j,k}_G(L')\,dL'\right]}=e^{-1}}$ then yields the
following condition for observability of the proposed instability in quantum superpositions:
\begin{equation}
\!\!\!\!\!\!L\sim\frac{l_{\!_P}}{8\pi\xi}\!\!
\left[\!\frac{3\left(m_j+m_k\right)}{5\,m^3_{\!_P}\,G^{\,}_{\!F}}-
\frac{m_j\,m_k\,E}{2\pi\,m^3_{\!_P}}
\ln\!\left\{\!\frac{6\pi e\left(m_j+m_k\right)}{5\,G^{\,}_{\!F}\,m_j\,m_k\,E}\!\!\right\}
\!\right]^{-1}\!\!\!\!\!\!\!\!,\label{time-length}
\end{equation}
where, for convenience, we have explicated the units by means of Planck length (${l_{\!_P}}$)
and Planck mass (${m_{\!_P}}$). 

From this observability condition it is easy to work out that---assuming the values of masses 
${m_{j}\approx m_{k}\approx 2}$ eV \cite{Kraus-2004}---the Di\'osi-Penrose scheme for state
reduction can be either ruled out or verified for the values of ${E}$ and ${L}$ in the
(approximate) ranges of ${[0,\,2.3]\times 10^{23}}$ eV and ${[0.7,\,15]\times 10^{9}}$
light-years,
respectively. As a result, provided cosmogenic neutrinos are at our disposal, an upper bound of
order ${10^{-2}}$ can be comfortably placed on the free parameter ${\xi}$. In fact, it may even
be possible to place an upper bound as strong as of order ${10^{-3}}$ on this parameter. This is
clear from the nature of the transition probability (\ref{modi-prob}) itself, which would change
significantly (thereby altering the observable flavor ratios from the quantum mechanical
expectations) even when the non-unitary Di\'osi-Penrose damping is as weak as ${e^{-0.1}}$.

{\em Observability of the non-unitary flavor oscillations}.---Ultra-high-energy neutrinos from
cosmologically distant sources such as active galactic nuclei (AGN) and gamma ray bursters (GRBs)
are generally believed to be produced as secondaries of cosmic ray protons interacting with
ambient matter and photon fields \cite{Athar-2002}. Such proton-proton and proton-photon
interactions produce neutral and charged pions, which, in turn, decay into neutrinos via the
chain:
${\pi^+\rightarrow\mu^+\nu_{\mu}\rightarrow e^+\,\nu_e\,\overline{\nu}_{\mu}\,\nu_{\mu}\,}$.
From the very inception, these interactions have been thought to provide a ``guaranteed'' source
of cosmogenic neutrinos. Moreover, although the absolute flux of the
different flavor states of such neutrinos is presently unknown, the above decay chain strongly
suggests their relative flux ratios
${\phi^S_{\nu_e}\!:\phi^S_{\nu_{\mu}}\!:\phi^S_{\nu_{\tau}}}$ at the source to be
${\frac{1}{3}:\frac{2}{3}:\frac{0}{3}}$.

Now, neutrinos---being stable and neutral particles---point back to their sources, thereby
providing vital information about their propagation lengths ${L}$, which can then be further
consolidated by the coincident data on the cosmological redshifts of the
sources \cite{Lunardini-2001}. Furthermore, being only
weakly interacting, in the absence of the Di\'osi-Penrose decay [provided condition
(\ref{obscon}) is satisfied] their flavor states (\ref{super-time}) would maintain quantum
coherence while propagating through the cosmic vacuum. Given the above initial flux ratios of
neutrino flavors, this coherence would then be reflected in the flavor fluxes observed at a
terrestrial detector, which can be easily calculated as
\begin{equation}
\phi^D_{\nu_{\beta}}(E,\,L)
\;=\!\!\sum_{\alpha\,=\,e,\,\mu,\,\tau}\!\!\!\!\!P_{\alpha\beta}(E,\,L)\;
\phi^S_{\nu_{\alpha}}\,,\label{at-detector}
\end{equation}
where the transition probabilities ${P_{\alpha\beta}}$ are given by (\ref{transi-prob}). The
corresponding flux ratios ${\phi^D_{\nu_e}\!:\phi^D_{\nu_{\mu}}\!:\phi^D_{\nu_{\tau}}}$ observed
at a detector can thus be compared with those predicted via the non-unitary transition
probabilities (\ref{modi-prob}), provided the mass-dependent observability relations
(\ref{time-length}) between the variables $E$ and $L$ in (\ref{at-detector}) are satisfied, and
sizable fluxes of relevant flavors are collected at Earth.

Fortunately, recent estimates of cosmogenic neutrino fluxes suggest that this is indeed
feasible. For example, the authors of Ref.\cite{Kalashev-2002} estimate sizable fluxes of
neutrinos from cosmologically distant sources in the energy ranges up to and beyond the
threshold of ${10^{21}}$ eV. In fact, a few neutrinos in the MeV range from
a distant supernova have already been observed \cite{Bionta-1987}. What is more, there are a
large variety of neutrino detectors under construction at present, or planned to be operational
in the near future, designed to be sensitive to a wide range of neutrino energies
\cite{Kalashev-2002}\cite{telescope}\cite{Christian-2005}. Their ability to measure the flavor
ratios of cosmogenic neutrinos at high precision has also been demonstrated in
Ref.\cite{Beacom-2003}. Therefore, a test of the
Di\'osi-Penrose ansatz by means of observing flavor ratios of cosmogenic neutrinos appears
to be quite feasible.

I am grateful to Roger Penrose for discussions on his ideas about gravity-induced
state reduction, and to Lajos Di\'osi for his constructive comments on the manuscript.

\end{document}